\documentclass[10pt,conference]{IEEEtran}

\IEEEoverridecommandlockouts

\usepackage{geometry}
\usepackage{hyperref}

\usepackage{ulem}
\usepackage{utfsym}
\usepackage{pifont}
\usepackage{booktabs}
\usepackage[shortlabels]{enumitem}
\usepackage{graphicx}
\usepackage{tabularx}
\usepackage{amsmath, amsthm}
\usepackage{amssymb}
\usepackage{mathtools}
\usepackage{comment}
\usepackage[subrefformat=parens,labelformat=parens]{subfig}
\usepackage{bm}
\usepackage{multirow}
\usepackage{threeparttable,booktabs}
\usepackage{tikz}
\usepackage[mathcal]{eucal}
\usepackage[]{algpseudocode}                               
\usepackage[ruled,lined,linesnumbered,noend]{algorithm2e}
\usepackage{filecontents}                                  
\usepackage{pgfplots}
\usepackage{pgfplotstable}
\pgfplotsset{compat=newest}
\usetikzlibrary{calc}
\usetikzlibrary{pgfplots.groupplots}

\theoremstyle{plain}

\newtheorem*{theorem*}{\textbf{Theorem}}

\theoremstyle{definition}

\newtheorem*{problem*}{\textbf{Problem}}

\usepackage[skip=1pt]{caption}            
\graphicspath{{./figs/}}

\definecolor{green(html/cssgreen)}{rgb}{0.0, 0.5, 0.0}

\begin{document}

\pagestyle{plain} 

\title{
  TimeTravelContainer: A Live Forking Sandbox\\for Stateful Agent Exploration
}

\author{
  \normalsize
  Zizheng Guo,
  Yanqing Zhang,
  Nathaniel Pinckney,
  Cunxi Yu,
  Brucek Khailany\\
  NVIDIA Corporation\\
  {\tt \{zizhengg,yanqingz,cunxiy,bkhailany\}@nvidia.com}
}

\maketitle
\newcommand\blfootnote[1]{%
  \begingroup
  \renewcommand\thefootnote{}\footnote{#1}%
  \addtocounter{footnote}{-1}%
  \endgroup
}

\begin{abstract}
LLM agents that drive stateful tools (e.g., interactive shells, design
environments, simulation engines) frequently need to clone, backtrack,
and parallelize live sessions to support exploration.
However, cloning a live session today requires serializing
the tool's in-memory state to disk or relaunching from scratch, both
prohibitively slow for interactive exploration.
Existing agent sandboxes
based on microVMs or kernel patches either serialize full guest memory to
storage or require privileged deployment, limiting use on shared compute
farms, personal computers, and other restricted environments.
We present TimeTravelContainer
(TTC), a rootless container runtime that live-forks an entire running
container, including processes, memory, filesystem, and network, with copy-on-write
(CoW) semantics.  Because the bulk of a tool session's memory is
typically read-only after the fork point, clones share it automatically
and only diverged pages consume additional memory.  TTC requires no
kernel modifications, hardware virtualization, or root privileges, and
runs on all major platforms including Linux, macOS (Apple Silicon), and Windows (WSL~2), making it
deployable as a portable foundation for agent harnesses.
We evaluate TTC on two agentic electronic design automation
(EDA) flows as representative stateful-tool workloads: MAGE
(Innovus) and EvoDRC (KLayout).  TTC achieves $84.1\times$ environment
cloning speedup, $31$--$72$\% memory reduction, and $1.7$--$9.4\times$
end-to-end speedup, demonstrating that live forking removes a scaling
barrier for agent-driven exploration of stateful tool environments.
\url{https://github.com/NVlabs/TimeTravelContainer}
\end{abstract}


\section{Introduction}
\label{sec:Introduction}

LLM agents are increasingly capable of driving complex, stateful tools
through multi-round interaction.  In software engineering, agents open
editors and terminals, propose code changes, run tests, and iterate based
on feedback~\cite{AGENT_sweagent, AGENT_openhands}.  In electronic design
automation (EDA), agents drive interactive tool shells to optimize a circuit~\cite{AGENT_chateda, AGENT_mage, AGENT_evodrc}.  Similar
patterns arise in robotics simulation, database tuning, scientific
computing, and anywhere an agent maintains a long-lived session with an
interactive tool.

A key capability for such agents is \emph{stateful exploration}: the
ability to clone a live tool session, try alternative actions in
parallel, and roll back to earlier states without losing work.  This
maps directly to tree search, Monte Carlo rollouts, and speculative
execution whose effectiveness depends on the cost of
creating and managing alternative execution states.  For agents to
achieve superhuman turnaround time and quality, the sandbox that hosts
the tool session must support instant cloning, cheap branching, and
transparent isolation.


Today's agent harnesses do not support this well.  Cloning a tool
session currently means either checkpointing the tool's state to disk and
restoring it, or relaunching the tool from scratch on a saved snapshot.
Existing agent sandboxes~\cite{TOOL_docker, TOOL_e2b} provide isolation
but no mechanism for cloning a live session.  MicroVM-based
platforms~\cite{TOOL_agentenv, TOOL_cubesandbox, TOOL_forkd} support VM
snapshots but serialize the full guest memory image to storage, which is
impractical when tools maintain tens of gigabytes of resident RAM.
Kernel-level approaches~\cite{SANDBOX_tclone, SANDBOX_deltabox}
achieve faster forking but require custom kernel modules or elevated
privileges, making them impractical on shared compute farms, personal laptops, and
edge devices where installing a kernel module is either complex and
insecure, or may require weeks of IT
approval if it is permitted at all. Since the sandbox will be
invoked by autorun agents directly, the additional exposure to external
malicious attacks from using non-standard kernel modules is also a
concern.
The lack of live forking support forces agents to either explore
sequentially, wasting the parallelism that would accelerate
search, or pay a steep latency and memory tax for each clone.

We present \emph{TimeTravelContainer} (TTC), a rootless Linux container
runtime with a \emph{live forking} primitive.  Built on
gVisor's~\cite{TOOL_gvisor} userspace kernel, TTC clones an entire running
container, including all processes, threads, memory, filesystem, and internal network
state, with copy-on-write (CoW) semantics.  Because all OS state is managed
in userspace, TTC implements container-level fork
without kernel patches, hardware virtualization, or root privileges, and can
itself run inside another sandbox.  TTC runs natively on Linux, Apple
Silicon Macs (via Hypervisor.framework), and Windows (through WSL~2),
making it a portable foundation for agent harnesses across platforms---from
cloud clusters to personal computers.
A fork creates independent clones that
share the same causal prefix: tool state that predates the fork point
remains shared; only memory pages that a clone actually modifies are
allocated.  This makes fork cost independent of resident memory size and
proportional only to the subsequent divergence.

We evaluate TTC on two agentic EDA flows as representative
stateful-tool workloads.  EDA is a stress test for live forking:
tools maintain tens to hundreds of gigabytes of resident RAM dominated by
read-only data structures, run as proprietary binaries that cannot be
modified, and execute on shared compute farms where root access is
typically unavailable.  We integrate TTC into
MAGE~\cite{AGENT_mage}, which drives Cadence Innovus for
interactive macro placement, and EvoDRC~\cite{AGENT_evodrc}, which uses
KLayout for evolutionary design-rule-violation repair.  Our contributions are:

\begin{enumerate}[leftmargin=*,nosep]
\item \textbf{A portable live-forking primitive for agent sandboxes.}
  TTC provides in-memory CoW container forking as a rootless, unprivileged
  runtime that requires no kernel modifications and runs on Linux, macOS,
  and Windows, addressing a performance gap in current agent infrastructure
  (Table~\ref{tab:sandbox}).

\item \textbf{Integration into two agentic circuit design flows.}
  We show how TTC replaces file-based checkpointing in MAGE/Innovus and
  tool relaunching in EvoDRC/KLayout, enabling transparent cloning and
  parallel exploration with no changes to the tools themselves.

\item \textbf{Systematic evaluation.}
  On MAGE, TTC achieves $84.1\times$ speedup on environment cloning
  ($86.5$\,s $\to$ $1.03$\,s) and $31$--$72$\% resident memory reduction
  across concurrent instances.  On EvoDRC, TTC delivers $1.8$--$2.1\times$
  end-to-end speedup on the original small benchmarks and
  $8$--$9.4\times$ on larger designs, demonstrating that live forking
  removes a scaling barrier that previously limited the scope of
  agent-driven exploration.
\end{enumerate}


\section{Preliminary}
\label{sec:Preliminary}

\subsection{Stateful Exploration in Agent Workflows}
\label{sec:prelim:agentic}

LLM agents increasingly drive stateful, interactive tools across
domains.  In software engineering, agents open editors, terminals, and
debuggers to propose patches, run tests, and iterate on
feedback~\cite{AGENT_sweagent, AGENT_openhands}.
In electronic design automation (EDA), agents have been applied across
the chip design flow:
RTL generation and repair~\cite{AGENT_verilogcoder, AGENT_autochip, AGENT_rtlfixer},
verification and testbench
construction~\cite{AGENT_autobench, AGENT_correctbench, AGENT_assertllm},
logic synthesis optimization~\cite{AGENT_drills, AGENT_boils, AGENT_rtlrewriter},
physical implementation~\cite{AGENT_chateda, AGENT_orfsagent, AGENT_edaid},
design-rule violation repair~\cite{AGENT_evodrc, AGENT_drccoder},
and end-to-end ASIC flows~\cite{AGENT_asicagent}.
Similar stateful interaction patterns arise in robotics simulation,
database administration, scientific computing, and any domain where an
agent maintains a long-lived session with an interactive tool.

A common pattern across these domains is \emph{stateful exploration}: the
agent opens an interactive tool session, proposes incremental changes,
evaluates feedback, and decides to proceed, backtrack, or clone into
parallel alternatives.  The effectiveness of this exploration is bounded
by the cost of managing alternative execution states.  When backtracking
requires a file-based checkpoint and relaunching the tool
from scratch, the agent either explores fewer alternatives or pays a
large wall-clock and memory penalty.  This bottleneck motivates a
general-purpose \emph{live forking} primitive for agent sandboxes.

\subsection{Agent Sandboxes and Live Forking}
\label{sec:prelim:sandbox}

Agent harnesses execute tool calls inside sandboxed environments for
isolation and reproducibility.  Most sandboxes prioritize
\emph{security}, i.e. preventing the agent from escaping the host, instead of
\emph{execution flexibility} such as cloning and restoring live states.
Docker~\cite{TOOL_docker} and OCI containers provide filesystem and
process isolation but no mechanism for cloning a running container.
SWE-agent~\cite{AGENT_sweagent} and OpenHands~\cite{AGENT_openhands} wrap
agent sessions in Docker; forking requires stopping, committing, and
relaunching, which discards in-memory state.
E2B~\cite{TOOL_e2b} provides cloud sandboxes with filesystem snapshots that
serialize state to disk without capturing live process memory.

\textbf{Virtualization approaches.} A growing body of work targets execution-state flexibility.
AgentENV~\cite{TOOL_agentenv}, CubeSandbox~\cite{TOOL_cubesandbox}, and
forkd~\cite{TOOL_forkd} build on microVM virtualization
(KVM/Firecracker) and support snapshotting or forking virtual machines.
However, their fork primitives serialize the \emph{full guest memory image}
to storage: the VM state must be written out and read back to create a
clone.  Stateful tools routinely maintain gigabytes to tens of gigabytes
of resident RAM which is dominated by read-only data (loaded models, parsed
databases, cached indices) that need not be copied at all. This makes
disk-based VM snapshots impractical for the sub-second cloning that
interactive agent exploration requires.

\textbf{Kernel-level approaches.}
Recent OS research addresses live forking directly.
TClone~\cite{SANDBOX_tclone} injects a kernel module into Linux to enable
cross-container copy-on-write (CoW) page-table sharing and extends
CRIU~\cite{TOOL_criu} to reconstruct process trees across namespaces
without serializing memory to disk, achieving $3.4\times$ lower fork
latency than stock CRIU.
DeltaBox~\cite{SANDBOX_deltabox} proposes an incremental checkpoint
abstraction and accelerates rollback by forking from a frozen template
process.
Fork-Explore-Commit~\cite{SANDBOX_fec} introduces a FUSE-based CoW
filesystem that gives each agent clone an isolated workspace, though it
operates at the filesystem level and does not fork live process state.
These kernel-module or kernel-patching approaches achieve fast forking but
require deploying a custom kernel or privileged module, limiting
applicability on shared compute clusters, managed cloud environments,
personal computers, and mobile or edge devices where agents increasingly
run.

\textbf{Unix \texttt{fork()} and its limitations.}
The \texttt{fork()} system call~\cite{OS_cow} creates a child process
sharing the parent's address space via CoW, making it nearly instantaneous
regardless of memory size.  However, \texttt{fork()} cannot directly clone
a complex application environment: it duplicates only the calling thread,
leaving all other threads terminated in the child; it does not preserve
process IDs, parent--child relationships, or session/group membership; and
it does not replicate open file descriptors pointing to container-local
resources such as internal sockets or pseudo-terminals.  A multi-threaded
tool with background threads, IPC channels, and a multi-process
architecture will crash or deadlock if merely injected with a
\texttt{fork()} call.

CRIU (Checkpoint/Restore in Userspace)~\cite{TOOL_criu} solves the
multi-process reconstruction problem by capturing the full state of a
process tree, including threads, PIDs, file descriptors, signal handlers, and
namespace membership. However, it does so by \emph{serializing the entire
memory image to disk}.  Restoration requires elevated privileges
(\texttt{CAP\_CHECKPOINT\_RESTORE} or \texttt{CAP\_SYS\_ADMIN}), and the
restored copy materializes every page independently, including the large
read-only regions that could have been shared, with no CoW sharing
with the original.

\textbf{gVisor and TTC.}
gVisor~\cite{TOOL_gvisor} reimplements the Linux kernel interface entirely
in userspace as a sandboxed process (the ``Sentry''), intercepting
application system calls without kernel modifications or hardware
virtualization.  Because the Sentry manages all OS state including page tables,
file descriptors, PIDs, and sockets in its own address space, it can
present to the guest processes a complete Linux kernel while implementing
its own layers of abstraction for security purposes.
We found the abstraction power of gVisor to extend beyond sandbox security
needs, and can serve as a general memory and OS resources virtualization
layer.
TimeTravelContainer (TTC) extends gVisor with exactly the primitives
agents need: a \emph{container-level fork} that clones the complete
environment (processes, threads, filesystem, network, memory) with CoW
semantics, without requiring kernel patches, KVM, or elevated privileges.
The technical details are elaborated in Section~\ref{sec:Algorithm}.
Table~\ref{tab:sandbox} compares TTC against existing platforms.

\begin{table}[t]
\centering
\caption{Comparison of agent execution environments.}
\label{tab:sandbox}
\vspace{-2mm}
\footnotesize
\setlength{\tabcolsep}{2.5pt}
\begin{tabular}{lcccc}
\toprule
\textbf{Platform} & \textbf{Live fork} & \textbf{In-mem} & \textbf{Unmodified} & \textbf{Unprivi-} \\
                  & \textbf{support}   & \textbf{CoW}    & \textbf{kernel} & \textbf{leged} \\
\midrule
Docker~\cite{TOOL_docker}              & \ding{55} & \ding{55} & \ding{51} & \ding{55} \\
SWE-agent~\cite{AGENT_sweagent}        & \ding{55} & \ding{55} & \ding{51} & \ding{55} \\
OpenHands~\cite{AGENT_openhands}       & \ding{55} & \ding{55} & \ding{51} & \ding{55} \\
E2B~\cite{TOOL_e2b}                    & \ding{55} & \ding{55} & \ding{51} & \ding{51} \\
AgentENV~\cite{TOOL_agentenv}          & \ding{51} & \ding{55} & \ding{55} & \ding{55} \\
CubeSandbox~\cite{TOOL_cubesandbox}    & \ding{51} & \ding{55} & \ding{55} & \ding{55} \\
forkd~\cite{TOOL_forkd}               & \ding{51} & \ding{55} & \ding{55} & \ding{55} \\
TClone~\cite{SANDBOX_tclone}           & \ding{51} & \ding{51} & \ding{55} & \ding{55} \\
DeltaBox~\cite{SANDBOX_deltabox}       & \ding{51} & \ding{51} & \ding{55} & \ding{55} \\
FEC~\cite{SANDBOX_fec}                 & \ding{51}$^*$ & \ding{55} & \ding{51} & \ding{51} \\
gVisor~\cite{TOOL_gvisor}              & \ding{55} & \ding{55} & \ding{51} & \ding{51} \\
\textbf{TTC (ours)}                    & \ding{51} & \ding{51} & \ding{51} & \ding{51} \\
\bottomrule
\multicolumn{5}{l}{\scriptsize $^*$Filesystem-only; does not fork live process state.}
\end{tabular}
\vspace{-3mm}
\end{table}


\section{TimeTravelContainer}
\label{sec:Algorithm}

\begin{figure*}[t]
\centering
\includegraphics[width=.92\textwidth]{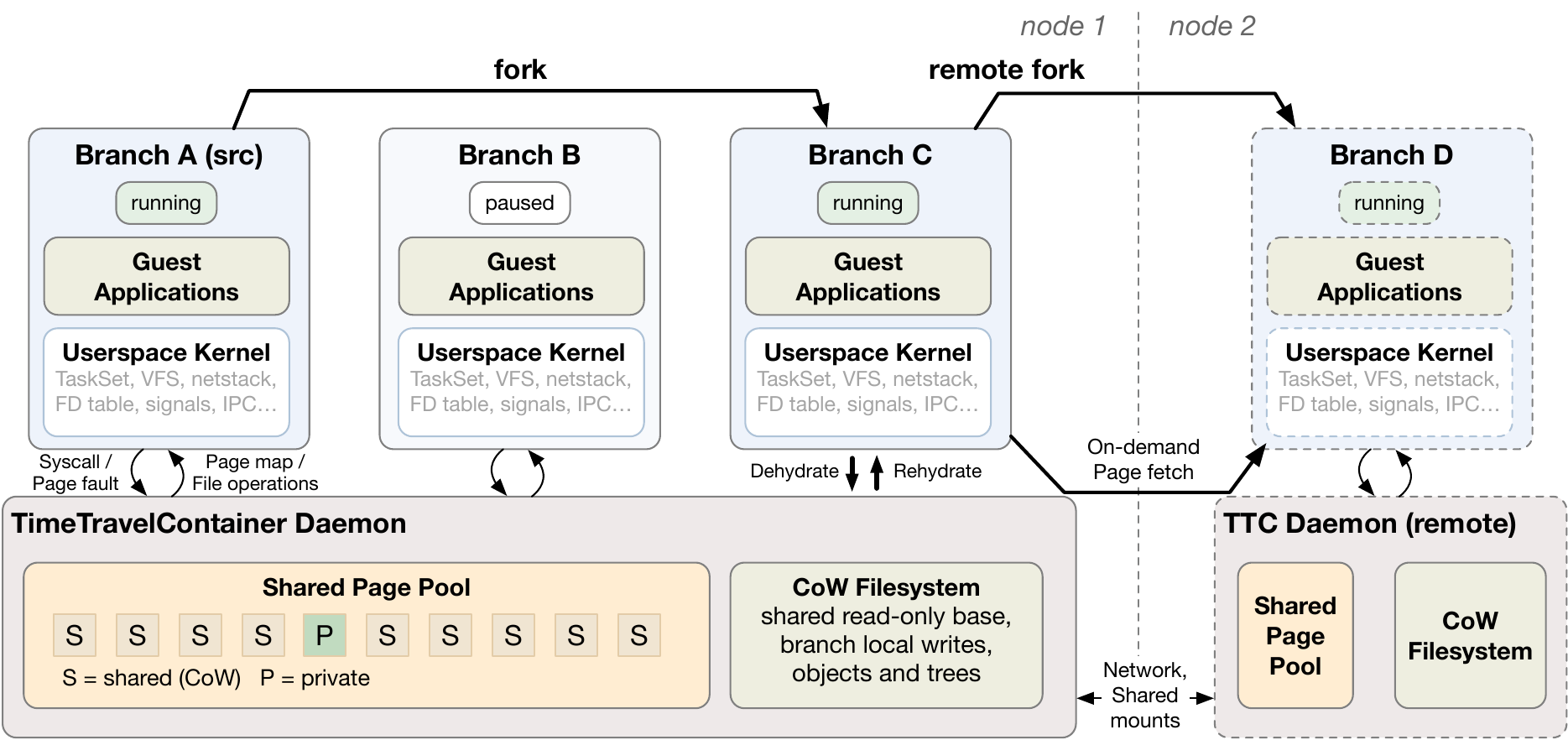}
\caption{TTC architecture.  All branches share one userspace platform and
  a common page pool.  Each branch maintains an independent kernel
  (processes, VFS, network).  Pages marked \textsf{S} are shared via CoW;
  pages marked \textsf{P} have diverged and are branch-private.  Operations
  shown: \textbf{fork} (instantaneous in-memory clone), \textbf{pause}
  (CPU release, memory resident), \textbf{dehydrate} (private pages to
  disk), and \textbf{remote fork} (cross-node clone with on-demand page
  fetch).}
\label{fig:arch}
\end{figure*}

Fig.~\ref{fig:arch} shows the architecture of TimeTravelContainer (TTC).
A sandbox family groups all clones of one container into Git-like branches.
The branches share a single platform instance (the daemon) and page pool but each owns an independent
kernel object graph with its own process tree, file descriptors, virtual
filesystem, and network stack.  The following subsections describe how TTC
realizes live forking, copy-on-write resource sharing, lifecycle
management, cross-platform portability, and cross-node scaling.

\subsection{A Userspace Kernel as Portable Foundation}
\label{sec:algo:foundation}

An agent harness must run in diverse environments: shared compute
clusters, managed clouds, personal laptops, and even inside other
sandboxes.  Requiring kernel modules, hardware virtualization,
or root privileges, as prior work
does~\cite{SANDBOX_tclone, SANDBOX_deltabox, TOOL_agentenv}, limits
deployment to environments where such modifications are permitted.

TTC adopts gVisor's userspace kernel~\cite{TOOL_gvisor} as its foundation.
The gVisor's userspace kernel (called \emph{Sentry}) is a single unprivileged process that intercepts
every system call and page fault from guest applications and simulates
the full Linux kernel interface: process management, virtual memory,
filesystems, networking, and signals.  Every guest process, whether a
multi-threaded application server, a database, or an EDA tool, sees a
complete, unmodified Linux environment.

Because all OS state lives inside the Sentry's address space rather than in
host kernel data structures, TTC can implement container-level fork,
pause, dehydration, and migration as operations on its own in-process data,
without any kernel cooperation.  These operations are invisible to the guest
application and require no co-design with it: the tool runs its unmodified
binary and is unaware that its environment has been forked or migrated.

\subsection{Copy-on-Write Memory for Parallel Exploration}
\label{sec:algo:mem}

Parallel exploration means running multiple copies of the same tool.
Naively, each copy materializes its own full memory image.  A single
interactive session may use gigabytes to tens of gigabytes of resident
memory. Running four parallel branches would multiply this proportionally,
quickly exhausting node capacity.

In practice, most of that memory does not change across branches: parsed
data structures, loaded libraries, cached indices, and tool state that
predates the fork point are all read-only.  TTC exploits this by
\emph{transparently sharing} identical pages across branches.

At fork time, TTC marks every memory page in the source branch as
copy-on-write: logically writable, but mapped read-only in hardware.
When a branch first writes to a shared page, the hardware MMU generates a
page fault.  TTC intercepts the fault, allocates a fresh 4\,KiB page,
copies the original content, and remaps the new page as writable.
Subsequent reads and writes to that page are handled
entirely by the hardware MMU with no TTC involvement.  The CoW retirement
cost is thus a one-time event per diverged page, and the steady-state
memory overhead is proportional only to the pages that \emph{actually
differ} between branches, not to the total resident size.


\subsection{Copy-on-Write Filesystem for Data Isolation}
\label{sec:algo:fs}

Stateful tool sessions often involve substantial on-disk data:
design files, model weights, database snapshots, intermediate results,
and configuration.  Agents and tools read and write these files, and
concurrent branches must not interfere with each other.

In software engineering agent harnesses such as SWE-agent~\cite{AGENT_sweagent}
and OpenHands~\cite{AGENT_openhands}, file isolation is typically achieved
via \texttt{git worktree}, which duplicates the entire working tree.  This
approach has two limitations for general stateful tools: it duplicates all
files regardless of whether they will be modified, consuming substantial
storage; and its strength lies in ASCII text files tracked by Git, while
many workloads predominantly use binary files or large artifacts that are
not Git-friendly.

TTC instead presents each branch with a virtual CoW file tree.  Every
file operation is intercepted at the userspace kernel level, where the Sentry
already owns the entire filesystem interface that guest processes interact
with.  Unmodified files are shared across branches by reference.  When a
branch first writes to a file, TTC creates a branch-local copy (using host
filesystem reflinks where available, or whole-file copy otherwise) and
redirects subsequent accesses to the local copy.  This requires no FUSE
module or kernel filesystem driver, because TTC already mediates all
guest I/O.

\subsection{Resource Management for Idle Branches}
\label{sec:algo:lifecycle}

Container branches frequently become idle.  A subagent may finish its
exploration, wait for LLM inference, or simply serve as a clean rollback
point that the main agent preserves in case subsequent attempts fail.
These idle branches still consume CPU time and memory if left running.

TTC provides two graduated mechanisms for releasing resources:

\textbf{Pause} stops all threads in the branch and releases CPU, but
keeps the branch's memory resident.  Resume is instantaneous: the threads
restart exactly where they stopped with no reloading.  This is suitable
for short idle periods such as waiting for LLM inference.

\textbf{Dehydrate} goes further: it writes the branch's \emph{private}
pages to persistent storage and releases the corresponding physical memory,
allowing a branch to become completely cold.  Two design decisions make
dehydration practical:

First, \emph{rehydration is instant from the tool's perspective}.  When a
dehydrated branch is resumed, TTC starts its processes immediately before
all memory pages have been loaded from disk.  Pages that the process
accesses are fetched on demand with priority; all remaining pages are loaded
asynchronously in the background.  The tool resumes execution on the first
instruction after dehydration without waiting for a bulk restore.

Second, \emph{dehydration preserves CoW relationships}.  Only pages that
are private to the dehydrated branch, i.e. pages that have already diverged
from other live branches, are saved.  Shared pages remain in the page pool,
referenced by the live branches that still need them.  When the branch is
later rehydrated, it re-establishes CoW references to the shared pages
rather than materializing independent copies.  A dehydrate--rehydrate cycle
therefore does not inflate memory usage.

\subsection{Cross-Platform Portability}
\label{sec:algo:portability}

Agent harnesses run in increasingly diverse settings: cloud clusters,
on-premises servers, developer laptops, and edge devices.  A live-forking
primitive tied to a single OS or architecture limits adoption.

TTC achieves cross-platform portability through its userspace kernel
design.  On \textbf{Linux hosts} (kernel $\geq$3.17), TTC intercepts system calls
via \texttt{seccomp}, without requiring hardware virtualization or nested virtualization at all. In environments where hardware virtualization is supported, an optional KVM backend is also available for lower syscall latency.
On \textbf{Apple Silicon Macs} (macOS 15.0+), TTC uses a
native Hypervisor.framework backend: it runs itself as the guest-facing
kernel inside the hypervisor, providing a full Linux environment without
booting a Linux VM.
On \textbf{Windows}, TTC runs through WSL~2.  In all
cases, the same container images, fork semantics, and CLI interface are fully supported.

TTC is an extensible platform in which users can add their own
abstractions of physical hardware and define their forking semantics.
As built-in examples, TTC supports in-container Vulkan and OpenGL graphics
acceleration and CUDA GPGPU on Linux, which allows booting a
forkable, graphics-accelerated Android~\cite{TOOL_redroid} or running GPU-accelerated
inference (e.g., llama.cpp with Vulkan and CUDA backends) inside a
forkable container.  This extensibility lets TTC adapt to workloads
that require specialized hardware, from EDA tools using compute
accelerators to agents driving graphical applications.

The special focus on portability enables TTC to serve as a foundation for agent
harnesses that must run across platforms: on shared EDA compute farms
where root access is unavailable, on a researcher's MacBook during
development, or on edge devices collecting agent trajectories for
continuous training data.

\subsection{Cross-Node Forking for Cluster-Scale Exploration}
\label{sec:algo:remote}

Stateful tools are resource-intensive, and production workloads typically
run on clusters of compute nodes.  When parallel exploration exceeds the
memory or CPU capacity of a single node, branches must spill to other
nodes.

TTC supports \emph{remote fork}: a branch on one node can be forked to
create a new branch on a different physical node, or an existing branch
can be migrated between nodes, all without interrupting the running processes.  The target
node starts the new branch immediately with an initially empty memory
image and fetches pages on demand: each 1\,MiB memory chunk is retrieved
from the source node or from shared storage (e.g., NFS or Lustre).

The filesystem layer follows the same pattern.  File objects from the shared
base image are accessible through the shared storage mount.  Branch-local
file modifications are transferred lazily alongside memory chunks.  This
design means that the wall-clock cost of a remote fork is dominated by the
pages the target actually touches, not by the total resident size of the
source branch.


\section{Experimental Results}
\label{sec:Results}

We evaluate TTC on two agentic electronic design automation (EDA)
flows as representative stateful-tool workloads.  EDA tools are a
demanding test case for live forking: they maintain tens to hundreds of
gigabytes of resident RAM, run as proprietary binaries that cannot be
instrumented, and operate on shared compute farms where root access is
unavailable.  We use MAGE~\cite{AGENT_mage}, where an LLM agent drives
Cadence Innovus for interactive macro placement; and
EvoDRC~\cite{AGENT_evodrc}, where an evolutionary agent drives KLayout for
DRC violation repair.  These two flows represent different integration
patterns: checkpoint/restore replacement (MAGE) and tool-relaunch
elimination (EvoDRC).

\textbf{Experimental setup.}
All experiments ran on servers with dual Intel Xeon Platinum 8270
processors (2.70\,GHz, 52 physical cores) and 755\,GiB RAM per node,
running Rocky Linux 8.9 with kernel 6.1.62.
Each tool operation used 8 worker threads, with aggregate concurrency
capped at 52 threads per node.
We used MAGE revision \texttt{27c3ed7}~\footnote{\url{https://github.com/ABKGroup/MAGE}}
with Cadence Innovus 23.37-s090\_1, and EvoDRC revision
\texttt{69a138b}~\footnote{\url{https://github.com/ASU-VDA-Lab/EvoDRC}}
with KLayout 0.30.12.
All experiments replay fixed tool workloads without invoking an LLM,
so reported times exclude LLM inference latency.

\subsection{Benchmarks}

Table~\ref{tab:benchmarks} summarizes the designs used in our experiments.
The MAGE benchmarks (Ariane133 and BlackParrot) are large-scale
designs from the MAGE flow.  The EvoDRC benchmarks comprise
seven small blocks (Block1--7) from the DAC'26 DRC benchmark
suite~\cite{BENCH_dac26drc} and three larger designs (AES, Ibex, JPEG)
that we adapted from ResizerAgent~\cite{AGENT_resizeragent}.

\begin{table}[t]
\centering
\caption{Benchmark designs.}
\label{tab:benchmarks}
\vspace{-2mm}
\footnotesize
\setlength{\tabcolsep}{4pt}
\begin{tabular}{llcrr}
\hline
\textbf{Flow} & \textbf{Design} & \textbf{Tech.} & \textbf{Cells} & \textbf{Nets} \\
\hline
MAGE       & Ariane133    & Nangate45 & 118\,594  & 172\,525  \\
MAGE       & BlackParrot  & Nangate45 & 719\,401  & 1\,354\,154 \\
\hline
DAC26 DRC     & Block1       & ASAP7     & 143       & 120       \\
DAC26 DRC     & Block2       & ASAP7     & 62        & 43        \\
DAC26 DRC     & Block3       & ASAP7     & 76        & 62        \\
DAC26 DRC     & Block4       & ASAP7     & 107       & 81        \\
DAC26 DRC     & Block5       & ASAP7     & 47        & 28        \\
DAC26 DRC     & Block6       & ASAP7     & 155       & 101       \\
DAC26 DRC     & Block7       & ASAP7     & 574       & 410       \\
\hline
ResizerAgent     & AES          & ASAP7     & 31\,163   & 18\,612   \\
ResizerAgent     & Ibex         & ASAP7     & 32\,705   & 23\,029   \\
ResizerAgent     & JPEG         & ASAP7     & 75\,803   & 61\,925   \\
\hline
\end{tabular}
\vspace{-2mm}
\end{table}

\subsection{MAGE: Cloning Speedup}

We compare two ways to create an independently usable copy of a
prepared Innovus session.  In the conventional MAGE workflow, the agent
saves an Innovus checkpoint to disk, launches a new Innovus process,
and restores the checkpoint; both saving and loading are I/O-bound and
scale with the on-disk image size.  With TTC, the agent instead forks
the running container, preserving the initialized Innovus process and
sharing its memory via copy-on-write.  Both paths complete after an
equivalent readiness query succeeds; we measure the total time from
initiating the clone to readiness.  These results isolate the cloning
primitive and do not include the MAGE agent's end-to-end runtime, which
is currently dominated by Innovus routing.

Table~\ref{tab:mage_clone} reports the results.  On Ariane133, TTC
reduces cloning latency from 86.5\,s to 1.03\,s, an $84.1\times$
speedup.  On the larger BlackParrot design (719\,K cells),
checkpoint/restore takes 185.7\,s; TTC fork completes in 3.0\,s,
a $61.8\times$ speedup.  TTC's fork cost is dominated by metadata
serialization and is largely independent of resident memory size.
After forking, each child independently performs connectivity checks,
DRC runs, and three edit-and-timing-report cycles to verify correct
isolated operation and to measure memory divergence under realistic
workloads (Table~\ref{tab:mage_mem}).

\begin{table}[t]
\centering
\caption{MAGE/Innovus: cloning latency comparison.}
\label{tab:mage_clone}
\vspace{-2mm}
\footnotesize
\setlength{\tabcolsep}{3.5pt}
\begin{tabular}{lccccc}
\hline
\textbf{Design} & \textbf{Save} & \textbf{Load} & \textbf{Total} & \textbf{TTC fork} & \textbf{Speed-} \\
                 & \textbf{(s)}  & \textbf{(s)}  & \textbf{(s)}   & \textbf{(s)}      & \textbf{up} \\
\hline
Ariane133   & 11.4  & 75.1  & 86.5  & 1.03 & $84.1\times$ \\
BlackParrot & 22.3  & 163.4 & 185.7 & 3.01 & $61.8\times$ \\
\hline
\end{tabular}
\vspace{-2mm}
\end{table}

\subsection{MAGE: Memory Reduction}

Table~\ref{tab:mage_mem} reports the total resident memory when running
multiple concurrent Innovus branches.  In the checkpoint/restore baseline,
each branch materializes an independent memory image, so total memory
scales linearly with the number of branches.  With TTC, branches share
read-only pages via copy-on-write: only pages that a branch actually
modifies are duplicated.

With 2 branches on Ariane133, TTC reduces memory by 37\% immediately
after forking (``Ready'').  After the branches diverge through connectivity
checks, DRC runs, and edit--timing-report cycles, TTC still saves 31\%
because most of the memory (netlist graph, libraries, and technology
data) remains shared.  With 6 branches the savings are more pronounced:
70\% at fork time, and 56\% after divergence.  The same trend holds on the
larger BlackParrot design, where 6 branches would require 62.4\,GB under
checkpoint/restore but only 40.9\,GB under TTC after full divergence, a
34\% reduction.

These results confirm that CoW sharing is effective for EDA workloads:
the bulk of resident memory consists of read-only data structures (the
netlist, libraries, and routing databases that predate the fork point),
and only the incremental modifications made by each branch consume
additional memory.

\begin{table}[t]
\centering
\caption{MAGE/Innovus: total resident memory (GB) across concurrent
  branches.  ``Saved'' is the fraction of memory eliminated by CoW sharing.}
\label{tab:mage_mem}
\vspace{-2mm}
\footnotesize
\setlength{\tabcolsep}{3pt}
\begin{tabular}{clrrr}
\hline
\textbf{Branches} & \textbf{Divergence step} & \textbf{Ckpt} & \textbf{TTC} & \textbf{Saved} \\
\hline
\multicolumn{5}{l}{\textit{Ariane133 (single instance: 5.0\,GB)}} \\
2 & Ready              &  9.1 &  5.8 & 37\% \\
  & Conn.\ + DRC       &  9.6 &  6.5 & 33\% \\
  & 3$\times$ edit+rpt & 9.9 &  6.8 & 31\% \\
6 & Ready              & 25.1 &  7.5 & 70\% \\
  & Conn.\ + DRC       & 27.7 & 12.3 & 56\% \\
  & 3$\times$ edit+rpt & 27.9 & 12.2 & 56\% \\
\hline
\multicolumn{5}{l}{\textit{BlackParrot (single instance: 8.8\,GB)}} \\
2 & Ready              & 17.1 & 10.6 & 38\% \\
  & Conn.\ + DRC       & 19.0 & 14.4 & 24\% \\
  & 3$\times$ edit+rpt & 20.5 & 15.5 & 24\% \\
6 & Ready              & 50.1 & 14.0 & 72\% \\
  & Conn.\ + DRC       & 59.7 & 36.1 & 40\% \\
  & 3$\times$ edit+rpt & 62.4 & 40.9 & 34\% \\
\hline
\end{tabular}
\vspace{-2mm}
\end{table}

\subsection{EvoDRC: End-to-End Tool Call Speedup}

In the conventional EvoDRC path, each candidate evaluation prepares
a Python layout context and launches a fresh KLayout process for DRC
checking.  With TTC, we prepare the context and a persistent KLayout
process once, then fork this environment for each candidate evaluation.
Each child inherits the fully initialized state while its modifications
remain isolated from the parent; candidate rendering and layout loading
remain part of the per-child evaluation.  We verified that geometry,
DRC reports, and repair verdicts match between the two paths.

Table~\ref{tab:evodrc} reports end-to-end runtime under two settings:
a single evolutionary call (one generation of mutations) and 10
consecutive calls (a full repair campaign).

On the small blocks (Block1--7), TTC delivers $1.8$--$2.3\times$ speedup
per call.  These blocks have modest KLayout startup cost, so the absolute
time saved per fork is small, but the relative improvement is consistent.
With 10 calls the speedups are similar ($1.8$--$2.1\times$), confirming
that the per-round savings compound linearly.

The advantage becomes dramatic on larger designs.  On AES (31\,K cells),
TTC achieves $31.9\times$ speedup on a single call and $8.0\times$ over 10
calls.  On Ibex and JPEG, the single-call speedups reach $103\times$ and
$173\times$, respectively.  These designs have large layout databases that
make KLayout relaunching extremely expensive; TTC's live fork bypasses this
cost entirely.  The 10-call speedups ($9.7\times$ and $9.4\times$) are
lower in ratio because they include the one-time initialization costs in the
beginning of sessions to demonstrate end-to-end speed-up, but the absolute time savings remain substantial:
JPEG drops from 23.7 hours to 2.5 hours.

These results demonstrate that TTC removes a scaling barrier in
agent-driven exploration.  On small designs where the tool's startup cost
is modest, TTC provides a steady ${\sim}2\times$ improvement.
The savings are significant for larger designs, turning what would be multi-day runs into
sub-hour experiments and making iterative agent exploration practical
on these cases for the first time.

\begin{table}[t]
\centering
\caption{EvoDRC/KLayout: end-to-end tool-call runtime comparison.}
\label{tab:evodrc}
\vspace{-2mm}
\footnotesize
\setlength{\tabcolsep}{2.5pt}
\begin{tabular}{lrrrrrr}
\hline
         & \multicolumn{3}{c}{\textbf{Single call}} & \multicolumn{3}{c}{\textbf{10 calls}} \\
\cmidrule(lr){2-4} \cmidrule(lr){5-7}
\textbf{Design} & \textbf{Orig.} & \textbf{TTC} & \textbf{Speedup}
                 & \textbf{Orig.} & \textbf{TTC} & \textbf{Speedup} \\
         & \textbf{(s)}   & \textbf{(s)}  & $\boldsymbol{\times}$
         & \textbf{(s)}   & \textbf{(s)}  & $\boldsymbol{\times}$\\
\hline
Block1 &     18.3 &    10.0 &   1.8 &     185 &    104 &  1.8 \\
Block2 &     13.5 &     7.1 &   1.9 &     136 &     72 &  1.9 \\
Block3 &     14.6 &     7.7 &   1.9 &     168 &     80 &  2.1 \\
Block4 &     16.9 &     9.1 &   1.9 &     169 &     92 &  1.8 \\
Block5 &     13.1 &     6.6 &   2.0 &     135 &     68 &  2.0 \\
Block6 &     18.9 &    10.6 &   1.8 &     191 &    112 &  1.7 \\
Block7 &     64.4 &    27.6 &   2.3 &     652 &    310 &  2.1 \\
\hline
AES    &  1\,318  &    41.3 &  31.9 & 13\,823 & 1\,725 &  8.0 \\
Ibex   &  3\,088  &    29.8 & 103.5 & 30\,879 & 3\,179 &  9.7 \\
JPEG   &  8\,527  &    49.3 & 173.0 & 85\,273 & 9\,032 &  9.4 \\
\hline
\end{tabular}
\vspace{-2mm}
\end{table}

\subsection{Discussion}

The MAGE and EvoDRC results demonstrate complementary aspects of TTC's
value.

\textbf{Cloning cost.}  MAGE shows that TTC reduces per-clone latency by
$61$--$84\times$ compared to Innovus checkpoint/restore.  This speedup is
the building block for any agent strategy that branches: tree search, Monte
Carlo rollouts, or speculative exploration.  Although MAGE's current
end-to-end runtime is dominated by Innovus routing (minutes per route), the
cloning speedup becomes decisive as agents move toward finer-grained
exploration with shorter per-branch work.

\textbf{Memory efficiency.}  CoW sharing reduces resident memory by
30--70\% depending on the degree of divergence.  This directly translates
to higher branch parallelism on a fixed memory budget: on Ariane133 with
6 branches, checkpoint/restore requires 25--28\,GB while TTC fits in
7.5--12.2\,GB, allowing either more branches or room for other workloads
on the same node.

\textbf{End-to-end impact.}  EvoDRC shows that when per-branch work is
short relative to tool startup cost, TTC's live fork delivers
transformative end-to-end speedups.  On larger designs where KLayout
loading dominates, TTC achieves $31$--$173\times$ single-call speedup.
TTC enables short per-branch exploration on a heavy tool state at scale, which
falls well in the advantage of LLM agents that prefer rapid trials with
immediate feedback.

\textbf{Generalizability beyond EDA.}
While our evaluation uses EDA tools, the properties that make live forking
effective here: large read-mostly memory footprints, expensive tool
initialization, and the need for isolated parallel exploration, are shared
by many stateful agent workloads.
Database agents that explore query plans or schema migrations on
multi-gigabyte databases face the same clone-or-relaunch dilemma.
Robotics simulation agents that evaluate action sequences in physics
engines would benefit from forking a simulator
mid-episode rather than replaying from the initial state.
Software engineering agents already use file-level isolation (git
worktree), but forking the entire environment with running servers,
databases, and build caches, would enable richer exploration.
TTC's CoW fork is tool-agnostic: it operates below the application layer,
requires no instrumentation or co-design with the guest tool, and its
benefits scale with the fraction of memory that remains shared across
branches.  Moreover, TTC's cross-platform portability (Linux, macOS,
Windows) and rootless deployment mean that these benefits are accessible
on personal computers and edge devices, opening
the possibility of collecting branched agent trajectories across
diverse real-world tasks for training continuously improving agentic
models.


\section{Conclusion}
\label{sec:Conclusion}

We presented TimeTravelContainer (TTC), a rootless, unprivileged
container runtime that live-forks running tool sessions with
copy-on-write semantics.  TTC requires no kernel modifications, hardware
virtualization, or root privileges, runs on all major platforms, making it a portable foundation for agent
harnesses across platforms.

Evaluated on two agentic EDA flows as representative stateful-tool
workloads, TTC achieves $84.1\times$ cloning speedup over Innovus
checkpoint/restore, 31--72\% memory reduction across concurrent
branches, and up to $9.4\times$ end-to-end speedup, demonstrating that
live forking removes a fundamental scaling barrier for agent-driven
exploration.

Beyond EDA, TTC enables a broader class of agent strategies that
combine multi-round interaction with parallel exploration.  Any domain
where agents drive stateful, long-lived tool sessions: software
engineering, database administration, robotics simulation, and scientific
computing, can benefit from instant, memory-efficient forking.
Because TTC's forking primitive is tool-agnostic and its deployment is
unprivileged, it extends to environments ranging from shared compute
clusters to personal computers and edge devices, opening the possibility
of collecting branched agent trajectories across diverse real-world tasks
for training continuously improving agentic models.


{
\small
\bibliographystyle{IEEEtran}
\bibliography{./ref/Top_sim,./ref/routing,./ref/ML,./ref/PD,./ref/cliche,./ref/DFM,./ref/Timing,./ref/ALG,./ref/Software,./ref/Simulation}
}

\end{document}